\documentclass[copyright,creativecommons]{eptcs}
\providecommand{\event}{MARS 2022} 
\usepackage{breakurl}              
\usepackage{underscore}            

\usepackage{cleveref}
\usepackage[normalem]{ulem}
\usepackage{wrapfig}
\usepackage{stmaryrd}
\usepackage{xspace}

\usepackage{tikz}
\usetikzlibrary{calc,automata,positioning,decorations.pathreplacing}
\tikzset{align at top/.style={baseline=(current bounding box.north)}}
\tikzstyle{every node}=[font=\scriptsize]
\tikzstyle{state} = [draw,fill=white,circle,thick,align=center,inner sep=0pt,minimum size=4.5mm]
\tikzstyle{lstate} = [draw,fill=white,rectangle,rounded corners,thick,align=center,inner sep=2pt]
\tikzstyle{dot} = [fill,circle,inner sep=0mm,minimum size=1.25mm,line width=0mm]

\usepackage{listings}
\makeatletter
\lst@AddToHook{OnEmptyLine}{\vspace{-0.75\baselineskip}}
\makeatother

\newcommand{\eg}{e.g.\ }
\newcommand{\ie}{i.e.\ }
\newcommand{\wrt}{w.r.t.\ }

\title{An Overview of Modest Models and Tools\\for Real Stochastic Timed Systems}
\author{Arnd Hartmanns\thanks{The author was supported by
NWO VENI grant no.\ 639.021.754
and by the European Union's Horizon 2020 research and innovation programme under the Marie Sk{\l}odowska-Curie grant agreement No.\ 101008233.}
\institute{University of Twente\\Enschede, The Netherlands}
\email{a.hartmanns@utwente.nl}
}
\def\titlerunning{Modest Models and Tools for Real Stochastic Timed Systems}
\def\authorrunning{A.\ Hartmanns}
\begin{document}
\maketitle

\begin{abstract}
We depend on the safe, reliable, and timely operation of cyber-physical systems ranging from smart grids to avionics components.
Many of them involve time-dependent behaviours and are subject to randomness.
Modelling languages and verification tools thus need to support these quantitative aspects.
In my invited presentation at MARS 2022, I gave an introduction to quantitative verification using the Modest modelling language and the Modest Toolset, and highlighted three recent case studies with increasing demands on model expressiveness and tool capabilities:
A case of power supply noise in a network-on-chip modelled as a Markov chain; a case of message routing in satellite constellations that uses Markov decision processes with distributed information; and a case of optimising an attack on Bitcoin via Markov automata model checking.
This paper summarises the presentation.
\end{abstract}

\section{Introduction}

Cyber-physical systems consist of discrete (usually digital, often implemented in software) controllers interacting with a continuous physical environment.
Control is often networked, sometimes wirelessly.
Many cyber-physical systems are safety- or performance-critical, or economically vital.
We thus need to ensure that they operate as desired, which includes dependability requirements such as reliability assurances, availability levels, or response time guarantees.
Reliability and availability are stochastic timed properties:
the probability of avoiding unsafe behaviour within a certain time horizon, and the expected fraction of time that the system is ready to provide service, respectively.
The critical systems themselves are also typically subject to randomisation, for example due to random message loss in wireless communication or due to employing randomised algorithms, and they are timed systems dealing with \eg transmission delays and timeouts or faults occurring unpredictably over time.
Thus, to assure their dependability by way of modelling and verification (ideally at design-time), we need stochastic timed formalisms and modelling languages supported by tools able to check stochastic timed properties.

In this overview paper accompanying my invited presentation at the 5th Workshop on Models for Formal Analysis of Real Systems (MARS 2022), I outline two such modelling languages, Modest and JANI, and one such set of tools, the Modest Toolset (in \Cref{sec:Modest}).
I then briefly summarise how Modest and the Modest Toolset have been used
to study power supply noise in a two-by-two network-on-chip system by way of a discrete-time Markov chain (DTMC) model and probabilistic model checking with the \textsf{mcsta} tool (in \Cref{sec:NoC});
to find routes through sparse constellations of nanosatellites using an abstract Markov decision process (MDP)~\cite{Bel57,How60} model analysed with a statistical model checking approach that employs scheduler sampling under distributed information as implemented in the \textsf{modes} tool (in \Cref{sec:Space});
and to optimise an attack on the Bitcoin cryptocurrency system via a Markov automata (MA)~\cite{EHZ10} model that permits \textsf{mcsta} to synthesise the strategy that minimises the expected time to success or maximises the probability of success within a certain time bound (in \Cref{sec:Bitcoin}).

\section{Modest Languages and Tools}
\label{sec:Modest}

\begin{figure}[t]
\centering
\begin{tikzpicture}[yscale=1.03,baseline={([yshift={-1.175\ht\strutbox}]current bounding box.north)}]
\tikzstyle{every node}=[font=\normalsize]
\draw[gray] (0,0) -- (3,1);
\draw[fill=white,color=white] (1.5,0.5) circle (0.05);
\draw (0,1) -- (1.5,2);
\draw (1.5,1) -- (3.0,2);
\draw (0,2) -- (0.75,3);
\draw (3,2) -- (2.25,3);
\draw (0.0,0) -- (1.5,1);
\draw (1.5,2) -- (0.75,3);
\draw (1.5,2) -- (2.25,3);
\draw (0.75,3) -- (1.5,4);
\draw (2.25,3) -- (1.5,4);
\draw[gray] (3,0) -- (3,1);
\draw (3.25,0) to[out=0,in=0] (3.05,2);
\draw (1.5,0) node [fill=white] {\textbf{DTMC}} --
      (1.5,1) node [fill=white] {\textbf{MDP}} --
      (1.5,2) node [fill=white] {PTA};
\draw (3,0) node [fill=white] {CTMC};
\draw (2.9,1) node [fill=white] {\textcolor{gray}{CTMDP}};
\draw (3,2) node [fill=white] {\textbf{MA}};
\draw (2.25,3) node [fill=white] {STA};
\draw (1.5,4) node [fill=white] {SHA};
\draw (0,0) node [fill=white] {LTS} --
      (0,1) node [fill=white] {TA} --
      (0,2) node [fill=white] {HA};
\draw (0.75,3) node [fill=white] {PHA};
\draw (4.35,2.705) node [anchor=east] {\begin{tiny}\raisebox{0.5pt}{+\,}\end{tiny}\begin{scriptsize}\textit{continuous}\end{scriptsize}};
\draw (4.31,2.455) node [anchor=east] {\begin{scriptsize}~~\,\textit{probability}\end{scriptsize}};
\draw (0,1.65) node [anchor=east] {\begin{tiny}\raisebox{0.5pt}{+\,}\end{tiny}\begin{scriptsize}\textit{continuous}\end{scriptsize}};
\draw[overlay] (0,1.4) node [anchor=east] {\begin{scriptsize}~~\,\textit{dynamics}\end{scriptsize}};
\draw (0,0.645) node [anchor=east] {\begin{tiny}\raisebox{0.5pt}{+\,}\end{tiny}\begin{scriptsize}\textit{real\phantom{\,}}\end{scriptsize}};
\draw (0,0.395) node [anchor=east] {\begin{scriptsize}\textit{time}\end{scriptsize}};
\draw (0,-0.3) node [] {\begin{scriptsize}\textit{nondeter-}\end{scriptsize}};
\draw[overlay] (0,-0.57) node [] {\begin{scriptsize}\textit{\strut ministic}\end{scriptsize}};
\draw[overlay] (0,-0.825) node [] {\begin{scriptsize}\textit{\strut choices}\end{scriptsize}};
\draw (1.5,-0.3) node [] {\begin{scriptsize}\textit{discrete}\end{scriptsize}};
\draw[overlay] (1.5,-0.57) node [] {\begin{scriptsize}\textit{\strut probabilities}\end{scriptsize}};
\draw (3,-0.3) node [] {\begin{scriptsize}\textit{exponential}\end{scriptsize}};
\draw[overlay] (3,-0.57) node [] {\begin{scriptsize}\textit{\strut residence}\end{scriptsize}};
\draw[overlay] (3,-0.825) node [] {\begin{scriptsize}\textit{\strut times}\end{scriptsize}};
\draw[] (4.5,4) node [fill=white] {\emph{Key:}};
\end{tikzpicture}
\begin{minipage}[t]{0.55\textwidth}
\renewcommand{\arraystretch}{0.95}
\begin{tabular}[t]{ll}
SHA & stochastic hybrid automata~\cite{FHHWZ11}\\
PHA & probabilistic hybrid automata~\cite{Spr00}\\
STA & stochastic timed automata~\cite{BDHK06}\\
HA & hybrid automata~\cite{ACHH92}\\
PTA & probabilistic timed automata~\cite{KNSS02}\\
MA & Markov automata\\
TA & timed automata~\cite{AD94}\\
MDP & Markov decision processes\\
CTMDP & continuous-time Markov decision processes\\
LTS & labelled transition systems\\
DTMC~ & discrete-time Markov chains\\
CTMC~ & continuous-time Markov chains\\
\end{tabular}
\end{minipage}
\caption{The family tree of automata-based quantitative formalisms}
\label{fig:ModelFamilyTree}
\end{figure}
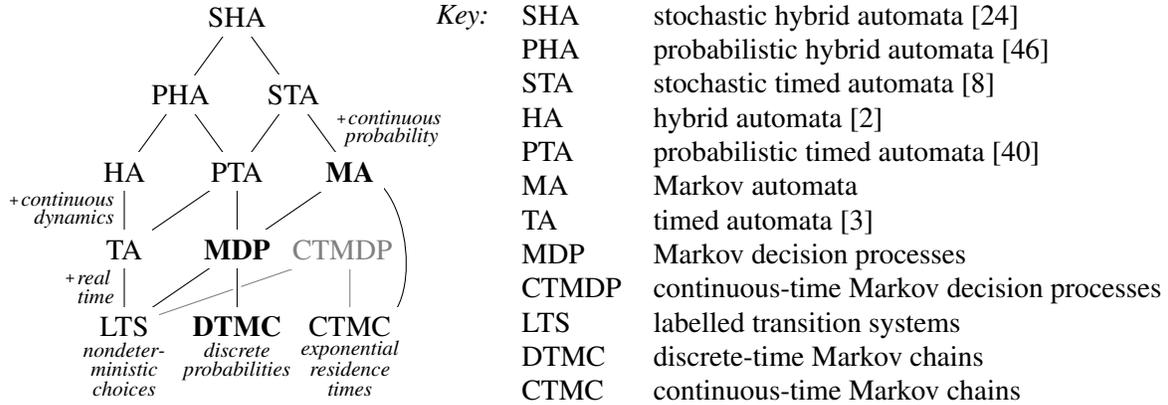 

A well-defined semantics in terms of some mathematically well-understood object is a cornerstone of formal models.
For quantitative models, we use automata-based formalisms---that represent the evolution of a system from state to state via (randomised) transitions---building on labelled transition systems (LTS, or Kripke structures) and discrete- and continuous-time Markov chains (DTMC and CTMC, respectively)~\cite{BK08}.
By combining these basic mathematical formalisms in various ways, and extending them with features such as real-time clocks and continuous variables evolving according to differential equations, we obtain further formalisms as depicted in \Cref{fig:ModelFamilyTree}.
Since writing real-life models as, say, large Markov chains would be cumbersome, we specify them using a higher-level modelling language that offers at least discrete variables with standard arithmetic and Boolean operators plus a notion of parallel composition for the natural specification of distributed and component- or actor-based systems.

\paragraph{The Modest Language.}
One such language is Modest, originally the \uline{mo}delling and \uline{de}scription language for \uline{s}tochastic \uline{t}imed systems~\cite{BDHK06}.
Its formal semantics was first defined in terms of STA and later extended to SHA~\cite{HHHK13}.
Modest is a textual modelling language; its syntax is designed to be similar to widely used programming languages like C or Java to lower the barrier of entry for domain experts.
At the same time, it is a process algebra in spirit, based on standard operators such as sequential and parallel composition, allowing the definition of and recursive calls to processes, and emphasising compositionality.
In fact, Modest consists of two largely orthogonal languages:
one to define \emph{behaviour}, which is the one based on process-algebraic ideas, and one to manipulate \emph{data} such as the values of discrete variables.
The latter provides arrays, recursive datatypes (\eg allowing the definition of a linked list type via pairs of a head containing a data item and a linked list option tail), and mutually recursive functions.
These features allow for concise and natural models of complex real-life systems.

\paragraph{The JANI model interchange format.}
While Modest is a convenient modelling language for end-users, the work required to implement code that parses Modest models and transforms the parsed syntax into its symbolic semantics (a parallel composition of SHA with discrete variables) is nontrivial.
The same problem affects many other modelling languages, \eg Prism's~\cite{KNP11}, too.
To ease tool development and facilitate the exchange of models between different tools, in 2016, the developers of several quantitative verification tools defined the JSON-based JANI~\cite{BDHHJT17} format.
It is not designed to be human-writable, but rather serve as a model interchange format that is generated by tools from other modelling languages, such as Modest.
Today, JANI is supported by the Modest Toolset (see below), Storm~\cite{DJKV17}, Momba~\cite{KKH21}, and several other tools.
All models in the quantitative verification benchmark set (QVBS)~\cite{HKPQR19} are available in both their original formats as well as in JANI.
The QVBS served as the foundation for the QComp 2019~\cite{HHHKKKPQRS19} and QComp 2020~\cite{BHKKPQTZ20} tool competitions.

\paragraph{The Modest Toolset.}
To support the creation of Modest models, and to compute the values of properties or check requirements specified as part of models, the Modest Toolset~\cite{HH14} provides a collection of visualisation, model transformation, model checking, and simulation tools.
The Modest Toolset has been in development since 2008; it is written in C\#, and is available as precompiled binaries for common Linux distributions, macOS, and Windows at \href{http://www.modestchecker.net/}{modestchecker.net}.
As input languages, it supports Modest and JANI; its \textsf{moconv} tool can convert between the two and apply various transformations, such as converting a suitable PTA model into its digital clocks~\cite{KNPS06} MDP.
The \textsf{mosta} tool visualises a model's symbolic semantics, helping in learning Modest and in debugging models.
The \textsf{mopy} tool converts a model into Python code implementing a first-state-next-state interface~\cite{BHKK03} that can be used to quickly prototype explicit-state verification algorithms and that is used by the author as part of the programming project of a Master's-level course on quantitative verification at the University of Twente.

The main implementation of probabilistic model checking (PMC)~\cite{Bai16,BAFK18} in the Modest Toolset is in \textsf{mcsta}~\cite{HH15}:
an explicit-state model checker that provides a unique disk-based approach to mitigate the state space explosion problem~\cite{HH15}.
It includes efficient model reductions such as the essential states abstraction~\cite{DJJL02}, and provides state-of-the-art algorithms for model checking MA~\cite{BHH21}.
The Modest Toolset's statistical model checker \textsf{modes}~\cite{BDHS20}complements \textsf{mcsta}'s capabilities for cases where model checking cannot be applied, such as when facing state space explosion or models with non-Markovian probability distributions like STA.
Statistical model checking (SMC)~\cite{AP18} is, in essence, Monte Carlo simulation applied to formal models and properties.
A constant-memory technique, it however incurs an explosion in runtime when faced with rare events, and does not directly support nondeterministic models such as MDP.
The \textsf{modes} tool addresses these shortcomings by providing rare event simulation~\cite{RT09} via a highly automated implementation of importance splitting~\cite{BDH19}, and by offering the lightweight scheduler sampling technique~\cite{LST14} for MDP, PTA~\cite{DHLS16,HSD17}, and (with limitations) stochastic-time models like MA and STA~\cite{DHS18}.
Other members of the Modest Toolset provide specialised analysis algorithms such as variants of the probabilistic planning algorithm LRTDP~\cite{BG03} for MDP in \textsf{modysh}~\cite{KH21} or an abstraction-based approach to safety verification of SHA in \textsf{prohver}~\cite{HHHK13}.

\section{Power Supply Noise in a Network-On-Chip System}
\label{sec:NoC}

As the complexity of distributed many-core systems advances, the network-on-chip (NoC) architecture has become the de-facto standard for on-chip communication.
A NoC is typically composed of topologically homogeneous routers operating synchronously in a decentralized manner using a predefined routing protocol.
Changes in the supply voltage---\emph{power supply noise} (PSN)---can influence the performance of the transistor devices in a NoC.
PSN is created by the simultaneous switching of logic devices, which causes a drop in the effective power supply voltage.
PSN is composed of two major components: resistive noise (related to the current drawn and the resistance of the circuit) and inductive noise (which is proportional to the rate of change of current through the inductance of the power grid).

\begin{wrapfigure}[10]{r}{7cm}
\vspace{-0.2cm}
\includegraphics[width=7cm]{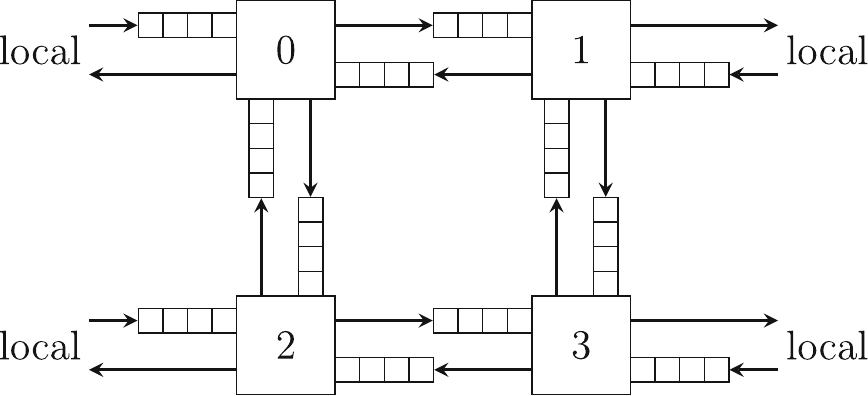}
\caption{Architecture of the $2 \times 2$ NoC~\cite{RLHBRCZ21}}
\label{fig:NocArch}
\end{wrapfigure}

To study PSN in NoC architectures, we modelled in Modest and analysed with \textsf{mcsta} first a single central router of a NoC~\cite{LHBSCRZ19} and later a two-by-two NoC consisting of four symmetric routers~\cite{RLHBRCZ21} as shown in \Cref{fig:NocArch}.
We focus on the latter in this section.
Our goal is to compute the probability for behavioural patterns that are likely to result in resistive resp.\ inductive noise to occur at least $n$ times within $t$ clock cycles, starting from an initial state where all buffers are empty.
We consider two different data packet (flit) generation patterns:
one where each router receives a flit into its local buffer (\eg from the one core it is connected to) every other cycle, and one where flits are generated in bursts.
We assume the destination of a flit to be one of the other router's local outputs, with the actual router selected uniformly at random for each flit.
The routers use a specific round-robin style routing protocol.
Thus, with all decisions fixed to be either deterministic (flit generation times and routing choices) or random (flit destinations), and the whole NoC running on a discrete clock, this system can naturally be modelled as a DTMC.
The main challenge for model checking with \textsf{mcsta} is to avoid the state space explosion problem.

For a first concrete model, which exploited the availability of complex user-defined datatypes in Modest to represent the state of the network's routers and buffers in full detail, we were unable to perform model checking for more than $t = 4$ clock cycles.
We then manually applied a series of abstractions to achieve tractability:
predicate abstraction to replace the details in the complex datatype's values by only the relevant predicates;
a probabilistic choice abstraction that delays random assignments to discrete variables to the point where the assigned value is first tested;
and an abstraction of the buffers that includes replacing them by bounded integer variables counting the number of waiting flits only.

\begin{wrapfigure}[16]{r}{7cm}
\vspace{-0.2cm}
\includegraphics[width=7cm]{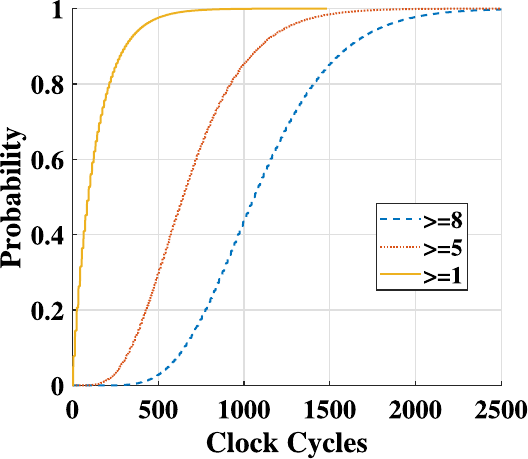}
\caption{CDF for inductive noise events~\cite{RLHBRCZ21}}
\label{fig:NocCdf}
\end{wrapfigure}

The resulting model could be model-checked for up to 30 clock cycles with every-other-cycle flit generation by unfolding the clock cycle counter into the state space, and up to any number of clock cycles by using the unfolding-free modified iteration technique of~\cite{HH16}, in essence computing the entire cumulative distribution function (CDF) as shown in \Cref{fig:NocCdf}.
This is due to an interesting effect of the different flit generation patterns:
With every-other-cycle generation, the buffers slowly fill up with flits to various destinations; the full state space that includes all combinations of buffer occupancies with different flits is too large to handle today.
Restricting the state space exploration to a bounded number of clock cycles, where initially few flits are present throughout the system, results in a sequence of manageable state spaces of ever-increasing size.
With bursty flit generation, all buffers periodically return to an empty state; the period is small enough for the entire state space to fit into memory, \ie buffers do not fill up far enough for the number of combinations of buffer states to grow too large, if clock cycles are managed as rewards.

We also applied SMC, which however was limited in the case of every-other-cycle generation by noise events being relatively rare, and in the case of bursty generation by not being able to compete in terms of runtime with the modified iteration technique that can compute the probabilities for the entire sequence of values of $t$ up to any upper bound in one go.
Similarly, our attempts to use Storm's binary decision diagram-based state space exploration did not provide scalability improvements, possibly due to the model not being as structured as we think it is, or simply due to a bad variable ordering in the model.
For further details on this first case study, we refer the interested reader to the original paper that was presented at FMICS~2021~\cite{RLHBRCZ21}.

\section{Routing in Satellite Constellations}
\label{sec:Space}

Satellite networks in low-Earth orbit are increasingly used to collect and distribute information across the globe, including access to the Internet.
For real-time applications like Internet access, this requires very large constellations (such as the Starlink constellation being deployed by SpaceX); even if low-cost satellites based on off-the-shelf components that are not space-qualified are used, the entire constellation becomes extremely expensive.
A different and more sustainable approach is to relax the real-time constraint and leverage the store-carry-and-forward principle where nodes store received messages for later forwarding to other nodes in the network, once a communication window---a contact---appears.
This gives rise to a delay-tolerant network.

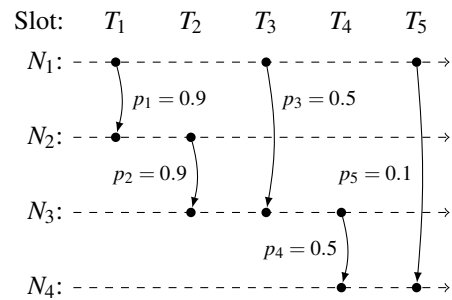
\begin{wrapfigure}[11]{r}{6.0cm}
\vspace{-0.4cm}
\centering
\begin{tikzpicture}[on grid,auto,align at top]
  \node[] (c00) [] {\small$N_1$:};
  \coordinate[right=5 of c00.east] (c50);
  \node[below=1 of c00] (c01) [] {\small$N_2$:};
  \coordinate[right=5 of c01.east] (c51);
  \node[below=1 of c01] (c02) [] {\small$N_3$:};
  \coordinate[right=5 of c02.east] (c52);
  \node[below=1 of c02] (c03) [] {\small$N_4$:};
  \coordinate[right=5 of c03.east] (c53);
  \node[dot] (n00) [right=0.5 of c00.east] {};
  \node[dot] (n01) [right=0.5 of c01.east] {};
  \node[dot] (n11) [right=1.5 of c01.east] {};
  \node[dot] (n12) [right=1.5 of c02.east] {};
  \node[dot] (n20) [right=2.5 of c00.east] {};
  \node[dot] (n22) [right=2.5 of c02.east] {};
  \node[dot] (n32) [right=3.5 of c02.east] {};
  \node[dot] (n33) [right=3.5 of c03.east] {};
  \node[dot] (n40) [right=4.5 of c00.east] {};
  \node[dot] (n43) [right=4.5 of c03.east] {};
  \node[] (t1) [above=0.25 of n00,anchor=south] {\small $T_1$};
  \node[] (t2) [right=1.0 of t1.south,anchor=south] {\small $T_2$};
  \node[] (t3) [right=1.0 of t2.south,anchor=south] {\small $T_3$};
  \node[] (t4) [right=1.0 of t3.south,anchor=south] {\small $T_4$};
  \node[] (t5) [right=1.0 of t4.south,anchor=south] {\small $T_5$};
  \node[overlay] (sl) [left=1.175 of t1.south,anchor=south] {\small \phantom{$T_1$}Slot:};
  ;
  \path[->]
    (c00) edge[dashed] node[] {} (c50)
    (c01) edge[dashed] node[] {} (c51)
    (c02) edge[dashed] node[] {} (c52)
    (c03) edge[dashed] node[] {} (c53)
  ;
  \path[-latex]
    (n00) edge[bend left=15] node[right] {$p_1=0.9$} (n01)
    (n11) edge[bend left=15] node[left] {$p_2=0.9$} (n12)
    (n20) edge[bend left=10] node[right,pos=0.22] {$p_3=0.5$} (n22)
    (n32) edge[bend left=15] node[left] {$p_4=0.5$} (n33)
    (n40) edge[bend left=5] node[left] {$p_5=0.1$} (n43)
  ;
\end{tikzpicture}%
\caption{Uncertain contact plan~\cite{DFH20}}
\label{fig:ContactPlan}
\end{wrapfigure}

In satellite constellations, the orbits are known with sufficient precision to calculate the upcoming contacts over the next few days, giving rise to a \emph{contact plan}.
However, message transmissions may fail for various reasons such as unreliable (low-cost) components, contact mispredictions, or interference during the wireless communication.
If statistical data is available or the error margins of calculations are known, we can assign a success probability to each contact, giving rise to an uncertain contact plan.
We show an abstract representation of such a plan in \Cref{fig:ContactPlan}.
This plan comprises four satellites (or ground stations) $N_1$ through $N_4$ with contacts over five time slots $T_1$ through $T_5$.
The numbers annotating contacts are the transmission success probabilities.

Now, given the source and destination of a message, and a limit $n$ on the number of message copies present in the network to avoid exhausting the satellites' limited resources, we would like to compute the routing strategy that maximises the probability of message delivery within the time window covered by the contact plan.
Due to the combination of randomness (in transmission failures) with nondeterministic decisions to be optimised (which contacts to use to send how many copies) in a discrete-time setting (a sequence of contacts), MDP are the perfect match among the formalisms of \Cref{fig:ModelFamilyTree} to model this problem.
The goal is to find an optimal scheduler (\ie routing strategy) for the MDP.

\begin{figure}[t]
\centering
\includegraphics[width=0.9\linewidth]{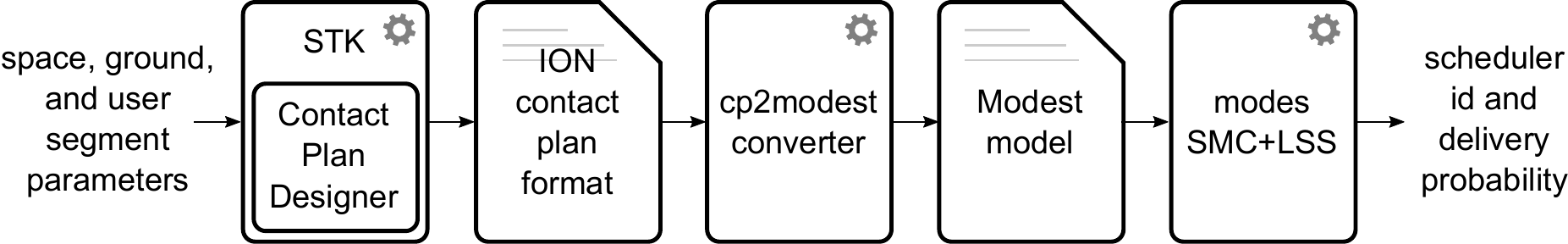}
\caption{Satellite routing scheduling toolchain for uncertain delay-tolerant networks~\cite{DFH20}}
\label{fig:SatelliteToolchain}
\end{figure}

We have tackled the problem by developing the toolchain outlined in \Cref{fig:SatelliteToolchain} that converts a concrete contact plan (with exact contact timings) into an abstract Modest MDP model.
At that point, one could apply PMC via \eg \textsf{mcsta} to obtain the optimal scheduler.
However, PMC works with complete, global information:
Consider the contact plan of \Cref{fig:ContactPlan} with $n = 2$.
$N_1$ will definitely send one copy to $N_2$ in slot $T_1$, and if successful, $N_2$ will forward that copy in slot $T_2$.
In slot $T_3$, the best course of action computed by PMC for satellite $N_1$ is to send its remaining copy to $N_3$ if any only if $N_3$ did not receive the first copy.
Thus the model checker ``sees'' the state of all satellites.
Satellites, however, do not have global information about the state of all other satellites in the constellation, making the optimal strategies found by PMC potentially unimplementable.
In fact, what we need are distributed schedulers~\cite{GD07}.
Unfortunately, the model checking problem under distributed schedulers is undecidable, and even with simplifications such as restricting to memoryless schedulers remains practically intractable~\cite{GD09}.
Recently, an approximative model checking-based approach that is specifically tailored to the uncertain delay-tolerant networks case has become available~\cite{RFMDFD21}, which however still remains limited by state space explosion as $n$ increases.

We instead propose to adapt the lightweight scheduler sampling (LSS) approach to sample distributed schedulers~\cite{DFH20}.
In LSS, each scheduler is represented by a fixed-size (\eg 32-bit) integer.
Performing an SMC analysis for each of $m$ randomly sampled such integers and keeping the maximum (minimum) estimate provides an underapproximation (overapproximation) for the maximum (minimum) probability achievable with the unknown optimal schedulers.
During an SMC analysis for scheduler $i$, when the simulator needs to decide between $k$ actions in state $s$, it concatenates the bitstring representations of $s$ and $i$, applies a hash function mapping this value to a fixed-size integer $j$, and selects the \mbox{$((j \mathbin{\,\mathrm{mod}\,} k) + 1)$-th} action.
To perform the same analysis \wrt distributed schedulers, all we need to change is the input to the hash function:
instead of the bitstring for $s$, we use that for a projection of $s$ to the variables observable by the currently active component (here: satellite).
We also introduce a condition of good-for-distribution models that, when satisfied, ensures that no two components may have a decision at the same time instant, making a global arbiter to break such ties unnecessary.
Our Modest models generated for uncertain contact plans are good for distribution by construction.

We implemented LSS for distributed schedulers as described above in \textsf{modes}, and applied this implementation to a small example contact plan as well as a realistic Walker-formation constellation.
Our results, presented at NFM 2020~\cite{DFH20}, show that LSS is able to find good and implementable routing strategies, and that restricting to distributed schedulers may actually result in better strategies than sampling from all (global-information) schedulers by virtue of restricting the sampling space.

\section{Optimally Attacking Bitcoin}
\label{sec:Bitcoin}

The Bitcoin cryptocurrency records its transactions in a blockchain to which blocks are added via the proof-of-work principle:
participants need to solve a computationally intensive problem to be able to generate or \emph{mine} a valid block.
Generally, the first new valid block mined gets appended to the chain, and a certain number of Bitcoins is awarded to the participant that found the block as a reward.
However, as a distributed system spanning the globe via the Internet, Bitcoin has to deal with asynchrony:
If multiple participants find new blocks at roughly the same time, there are different alternative forks of the Bitcoin blockchain, and a consensus must be reached on which is the valid one.
In Bitcoin, the longest available chain is considered the valid one.

As the computational power used for mining new blocks (the hash rate) changes, the Bitcoin network periodically adjusts the hardness of the problem such that the average time to find a new block (the confirmation time) is 10 minutes.
In practice, the actual confirmation time varies; it was about 12 minutes in 2017~\cite{FC18}.
This time is truly random, and the mining of new blocks can abstractly be modelled by a CTMC in which the transition from a chain with $n$ blocks to one with $n+1$ blocks has rate $\frac{1}{12}$, \ie the time until the transition is taken is exponentially distributed with that rate.

If a large amount of the hash rate (say $M$ percent) is controlled by one malicious entity, they could feasibly implement various attacks on the Bitcoin network by secretly working on their own fork until it becomes longer than the ``public'' one, and then broadcasting the secret fork.
For example, Bitcoins could be spent twice:
once on the public fork in block $b_i$, and once on the secret fork that branches off from publicly known block $b_j$ that is before $b_i$ in the chain.
This behaviour can be integrated into an abstract CTMC model of Bitcoin to \eg compute the expected time until the attack succeeds for various values of $M$.
We built such a model in Modest and studied similar properties using \textsf{mcsta} and \textsf{modes}~\cite{HH19}.

\begin{figure}[t]
\begin{lstlisting}
const real M;              // fraction of hash rate controlled by malicious mining pool
const int CD;              // confirmation depth required by victim
const int DB = CD;         // attacker gives up when this far behind
\end{lstlisting}~\\[-30pt]
\begin{lstlisting}
action sln;                // indicates that the honest pool mined a new block
action rst;                // indicates that the attacker restarts from the public fork
action cnt;                // indicates that the attacker continues
\end{lstlisting}~\\[-30pt]
\begin{lstlisting}
int(0..CD+1) m_len;        // length of the secret fork
int(-DB..CD+1) m_diff = 0; // length of secret fork minus honest fork
\end{lstlisting}~\\[-30pt]
\begin{lstlisting}
process HonestPool()
{
   rate(1/12 * (1 - M)) tau; // wait 12 / (1 - M) minutes on average
   sln;                      // signal that a new block was found
   HonestPool()              // repeat
}
\end{lstlisting}~\\[-30pt]
\begin{lstlisting}
process TrustAttacker()
{
   do {
   :: rate((1/12) * M) {= m_len = min(CD, m_len + 1), m_diff++ =} // new secret block
   :: sln {= m_diff-- =};                // public fork extended
      alt {                              // strategy choice: restart or continue malicious fork
      :: rst {= m_len = 0, m_diff = 0 =} // can always restart
      :: when(m_diff > -DB) cnt          // can continue if not too far behind
      }
   }
}
\end{lstlisting}~\\[-30pt]
\begin{lstlisting}
par {
:: HonestPool()
:: TrustAttacker()
}
\end{lstlisting}
\caption{Modest model for optimising the trust attack on Bitcoin~\cite{HH19}}
\label{fig:BitcoinModest}
\end{figure}

A more interesting and somewhat easier attack, which however does not have the individual benefit of doubly-spent coins but rather attempts to undermine the public trust in Bitcoin, is to simply try to obtain a secret fork that is longer than the official one by a certain margin, and then publish that fork.
If done repeatedly, regular users could no longer rely on the persistence of transactions that initially appeared to have become a part of the valid Bitcoin blockchain.
In this attack, every time the public fork is extended, the malicious entity may decide between (a)~continuing to work on its current secret fork and (b)~restarting its secret fork from the new public block.
This is because it is no longer necessary to purge a specific block $b_i$ from the public chain as in the double-spending attack.
Due to the presence of a nondeterministic choice to be optimised, this attack can thus no longer be represented in a CTMC model.

The attack on trust in Bitcoin was first analysed by Fehnker and Chaudhary~\cite{FC18} using statistical model checking with UPPAAL SMC~\cite{DLLMW11}.
As a consequence of using SMC, they had to run a separate analysis for every possible strategy determining the conditions for when to continue and when to restart, and their results came with a statistical error.
We later modelled the same scenario as the Modest MA model shown in \Cref{fig:BitcoinModest} and let \textsf{mcsta} synthesise the optimal strategy~\cite{HH19}, which it could do in a matter of a few seconds.
We found that the optimal strategy is to restart the attack if
(i)~the public chain is extended when the secret fork is still empty,
(ii)~the secret fork has one block and the public fork adds a third new block, or
(iii)~the secret fork has $\geq 2$ blocks and becomes three blocks shorter than the public one,
and to continue the attack in all other cases.
If the malicious entity controls just 20\,\% of the hash rate, which is not an uncommon situation for the Bitcoin network, then the expected time to success under this strategy is only approximately 2.5 days.

\section{Summary}

Different case studies have different needs in terms of conceptual modelling power, modelling language features, and analysis tool capabilities.
I highlighted three examples that were modelled in the Modest language and analysed using different tools from the Modest Toolset:
First, in the case of \textbf{power supply noise in a NoC}, the simple formalism of DTMC was sufficient.
For the detailed concrete model, however, the Modest language feature of declaring and using one's own complex data types was very helpful.
PMC via \textsf{mcsta} was the analysis method of choice, however significant effort was needed to abstract the model until it became tractable for PMC due to the state space explosion problem.
Second, for \textbf{routing in satellite constellations}, nondeterministic choices needed to be modelled, and optimised over by the analysis tool.
Here, MDP fit the problem very well with their ability to model decision-making under uncertainty.
We auto-generated Modest models from contact plans computed by domain-specific software.
Due to the need to find implementable routing strategies in the distributed-information setting of satellite constellations, we could not use PMC; instead, we adapted the LSS approach to allow SMC to handle both distributed information and nondeterminism.
Finally, to \textbf{optimally attack Bitcoin}, we showed that MA fit the problem well due to the combination of the stochastic time-to-next-block with the nondeterministic choices between continuing and restarting the secret fork.
Using PMC with \textsf{mcsta} again, we were able to compute an optimal strategy with little computational effort.

\paragraph{Acknowledgments.}
I thank my co-authors for the papers underlying the three case studies~\cite{DFH20,HH19,LHBSCRZ19,RLHBRCZ21}, without whom my presentation at MARS and this summary paper would not have been possible:
Prabal Basu, Koushik Chakraborty, Pedro R. D'Argenio, Juan A. Fraire, Holger Hermanns, Rajesh Jayashankara Shridevi, Benjamin Lewis, Riley Roberts, Sanghamitra Roy, and Zhen Zhang.

\bibliographystyle{eptcs}
\bibliography{abstract}

\begin{thebibliography}{10}
\providecommand{\bibitemdeclare}[2]{}
\providecommand{\surnamestart}{}
\providecommand{\surnameend}{}
\providecommand{\urlprefix}{Available at }
\providecommand{\url}[1]{\texttt{#1}}
\providecommand{\href}[2]{\texttt{#2}}
\providecommand{\urlalt}[2]{\href{#1}{#2}}
\providecommand{\doi}[1]{doi:\urlalt{http://dx.doi.org/#1}{#1}}
\providecommand{\eprint}[1]{arXiv:\urlalt{https://arxiv.org/abs/#1}{#1}}
\providecommand{\bibinfo}[2]{#2}

\bibitemdeclare{article}{AP18}
\bibitem{AP18}
\bibinfo{author}{Gul \surnamestart Agha\surnameend} \& \bibinfo{author}{Karl
  \surnamestart Palmskog\surnameend} (\bibinfo{year}{2018}):
  \emph{\bibinfo{title}{A Survey of Statistical Model Checking}}.
\newblock {\sl \bibinfo{journal}{{ACM} Trans. Model. Comput. Simul.}}
  \bibinfo{volume}{28}(\bibinfo{number}{1}), pp. \bibinfo{pages}{6:1--6:39},
  \doi{10.1145/3158668}.

\bibitemdeclare{inproceedings}{ACHH92}
\bibitem{ACHH92}
\bibinfo{author}{Rajeev \surnamestart Alur\surnameend}, \bibinfo{author}{Costas
  \surnamestart Courcoubetis\surnameend}, \bibinfo{author}{Thomas~A.
  \surnamestart Henzinger\surnameend} \& \bibinfo{author}{Pei-Hsin
  \surnamestart Ho\surnameend} (\bibinfo{year}{1992}):
  \emph{\bibinfo{title}{Hybrid Automata: An Algorithmic Approach to the
  Specification and Verification of Hybrid Systems}}.
\newblock In \bibinfo{editor}{Robert~L. \surnamestart Grossman\surnameend},
  \bibinfo{editor}{Anil \surnamestart Nerode\surnameend},
  \bibinfo{editor}{Anders~P. \surnamestart Ravn\surnameend} \&
  \bibinfo{editor}{Hans \surnamestart Rischel\surnameend}, editors: {\sl
  \bibinfo{booktitle}{Hybrid Systems}}, {\sl \bibinfo{series}{Lecture Notes in
  Computer Science}} \bibinfo{volume}{736}, \bibinfo{publisher}{Springer}, pp.
  \bibinfo{pages}{209--229}, \doi{10.1007/3-540-57318-6\_30}.

\bibitemdeclare{article}{AD94}
\bibitem{AD94}
\bibinfo{author}{Rajeev \surnamestart Alur\surnameend} \&
  \bibinfo{author}{David~L. \surnamestart Dill\surnameend}
  (\bibinfo{year}{1994}): \emph{\bibinfo{title}{A Theory of Timed Automata}}.
\newblock {\sl \bibinfo{journal}{Theor. Comput. Sci.}}
  \bibinfo{volume}{126}(\bibinfo{number}{2}), pp. \bibinfo{pages}{183--235},
  \doi{10.1016/0304-3975(94)90010-8}.

\bibitemdeclare{incollection}{Bai16}
\bibitem{Bai16}
\bibinfo{author}{Christel \surnamestart Baier\surnameend}
  (\bibinfo{year}{2016}): \emph{\bibinfo{title}{Probabilistic Model Checking}}.
\newblock In \bibinfo{editor}{Javier \surnamestart Esparza\surnameend},
  \bibinfo{editor}{Orna \surnamestart Grumberg\surnameend} \&
  \bibinfo{editor}{Salomon \surnamestart Sickert\surnameend}, editors: {\sl
  \bibinfo{booktitle}{Dependable Software Systems Engineering}}, {\sl
  \bibinfo{series}{{NATO} Science for Peace and Security Series -- {D}:
  Information and Communication Security}}~\bibinfo{volume}{45},
  \bibinfo{publisher}{{IOS} Press}, pp. \bibinfo{pages}{1--23},
  \doi{10.3233/978-1-61499-627-9-1}.

\bibitemdeclare{incollection}{BAFK18}
\bibitem{BAFK18}
\bibinfo{author}{Christel \surnamestart Baier\surnameend},
  \bibinfo{author}{Luca \surnamestart de~Alfaro\surnameend},
  \bibinfo{author}{Vojtech \surnamestart Forejt\surnameend} \&
  \bibinfo{author}{Marta \surnamestart Kwiatkowska\surnameend}
  (\bibinfo{year}{2018}): \emph{\bibinfo{title}{Model Checking Probabilistic
  Systems}}.
\newblock In \bibinfo{editor}{Edmund~M. \surnamestart Clarke\surnameend},
  \bibinfo{editor}{Thomas~A. \surnamestart Henzinger\surnameend},
  \bibinfo{editor}{Helmut \surnamestart Veith\surnameend} \&
  \bibinfo{editor}{Roderick \surnamestart Bloem\surnameend}, editors: {\sl
  \bibinfo{booktitle}{Handbook of Model Checking}},
  \bibinfo{publisher}{Springer}, pp. \bibinfo{pages}{963--999},
  \doi{10.1007/978-3-319-10575-8\_28}.

\bibitemdeclare{book}{BK08}
\bibitem{BK08}
\bibinfo{author}{Christel \surnamestart Baier\surnameend} \&
  \bibinfo{author}{Joost-Pieter \surnamestart Katoen\surnameend}
  (\bibinfo{year}{2008}): \emph{\bibinfo{title}{Principles of model checking}}.
\newblock \bibinfo{publisher}{{MIT} Press}.

\bibitemdeclare{article}{Bel57}
\bibitem{Bel57}
\bibinfo{author}{Richard \surnamestart Bellman\surnameend}
  (\bibinfo{year}{1957}): \emph{\bibinfo{title}{A {M}arkovian decision
  process}}.
\newblock {\sl \bibinfo{journal}{Journal of Mathematics and Mechanics}}
  \bibinfo{volume}{6}(\bibinfo{number}{5}), pp. \bibinfo{pages}{679--684}.

\bibitemdeclare{article}{BDHK06}
\bibitem{BDHK06}
\bibinfo{author}{Henrik~C. \surnamestart Bohnenkamp\surnameend},
  \bibinfo{author}{Pedro~R. \surnamestart D'Argenio\surnameend},
  \bibinfo{author}{Holger \surnamestart Hermanns\surnameend} \&
  \bibinfo{author}{Joost-Pieter \surnamestart Katoen\surnameend}
  (\bibinfo{year}{2006}): \emph{\bibinfo{title}{{MoDeST}: A Compositional
  Modeling Formalism for Hard and Softly Timed Systems}}.
\newblock {\sl \bibinfo{journal}{{IEEE} Trans. Software Eng.}}
  \bibinfo{volume}{32}(\bibinfo{number}{10}), pp. \bibinfo{pages}{812--830},
  \doi{10.1109/TSE.2006.104}.

\bibitemdeclare{inproceedings}{BHKK03}
\bibitem{BHKK03}
\bibinfo{author}{Henrik~C. \surnamestart Bohnenkamp\surnameend},
  \bibinfo{author}{Holger \surnamestart Hermanns\surnameend},
  \bibinfo{author}{Joost-Pieter \surnamestart Katoen\surnameend} \&
  \bibinfo{author}{Ric \surnamestart Klaren\surnameend} (\bibinfo{year}{2003}):
  \emph{\bibinfo{title}{The {M}odest Modeling Tool and Its Implementation}}.
\newblock In \bibinfo{editor}{Peter \surnamestart Kemper\surnameend} \&
  \bibinfo{editor}{William~H. \surnamestart Sanders\surnameend}, editors: {\sl
  \bibinfo{booktitle}{13th International Conference on Computer Performance
  Evaluations, Modelling Techniques and Tools ({TOOLS})}}, {\sl
  \bibinfo{series}{Lecture Notes in Computer Science}} \bibinfo{volume}{2794},
  \bibinfo{publisher}{Springer}, pp. \bibinfo{pages}{116--133},
  \doi{10.1007/978-3-540-45232-4\_8}.

\bibitemdeclare{inproceedings}{BG03}
\bibitem{BG03}
\bibinfo{author}{Blai \surnamestart Bonet\surnameend} \&
  \bibinfo{author}{Hector \surnamestart Geffner\surnameend}
  (\bibinfo{year}{2003}): \emph{\bibinfo{title}{Labeled {RTDP}: Improving the
  Convergence of Real-Time Dynamic Programming}}.
\newblock In \bibinfo{editor}{Enrico \surnamestart Giunchiglia\surnameend},
  \bibinfo{editor}{Nicola \surnamestart Muscettola\surnameend} \&
  \bibinfo{editor}{Dana~S. \surnamestart Nau\surnameend}, editors: {\sl
  \bibinfo{booktitle}{13th International Conference on Automated Planning and
  Scheduling ({ICAPS})}}, \bibinfo{publisher}{{AAAI}}, pp.
  \bibinfo{pages}{12--21}.

\bibitemdeclare{article}{BDH19}
\bibitem{BDH19}
\bibinfo{author}{Carlos~E. \surnamestart Budde\surnameend},
  \bibinfo{author}{Pedro~R. \surnamestart D'Argenio\surnameend} \&
  \bibinfo{author}{Arnd \surnamestart Hartmanns\surnameend}
  (\bibinfo{year}{2019}): \emph{\bibinfo{title}{Automated compositional
  importance splitting}}.
\newblock {\sl \bibinfo{journal}{Sci. Comput. Program.}} \bibinfo{volume}{174},
  pp. \bibinfo{pages}{90--108}, \doi{10.1016/j.scico.2019.01.006}.

\bibitemdeclare{article}{BDHS20}
\bibitem{BDHS20}
\bibinfo{author}{Carlos~E. \surnamestart Budde\surnameend},
  \bibinfo{author}{Pedro~R. \surnamestart D'Argenio\surnameend},
  \bibinfo{author}{Arnd \surnamestart Hartmanns\surnameend} \&
  \bibinfo{author}{Sean \surnamestart Sedwards\surnameend}
  (\bibinfo{year}{2020}): \emph{\bibinfo{title}{An efficient statistical model
  checker for nondeterminism and rare events}}.
\newblock {\sl \bibinfo{journal}{Int. J. Softw. Tools Technol. Transf.}}
  \bibinfo{volume}{22}(\bibinfo{number}{6}), pp. \bibinfo{pages}{759--780},
  \doi{10.1007/s10009-020-00563-2}.

\bibitemdeclare{inproceedings}{BDHHJT17}
\bibitem{BDHHJT17}
\bibinfo{author}{Carlos~E. \surnamestart Budde\surnameend},
  \bibinfo{author}{Christian \surnamestart Dehnert\surnameend},
  \bibinfo{author}{Ernst~Moritz \surnamestart Hahn\surnameend},
  \bibinfo{author}{Arnd \surnamestart Hartmanns\surnameend},
  \bibinfo{author}{Sebastian \surnamestart Junges\surnameend} \&
  \bibinfo{author}{Andrea \surnamestart Turrini\surnameend}
  (\bibinfo{year}{2017}): \emph{\bibinfo{title}{{JANI}: Quantitative Model and
  Tool Interaction}}.
\newblock In \bibinfo{editor}{Axel \surnamestart Legay\surnameend} \&
  \bibinfo{editor}{Tiziana \surnamestart Margaria\surnameend}, editors: {\sl
  \bibinfo{booktitle}{23rd International Conference on Tools and Algorithms for
  the Construction and Analysis of Systems ({TACAS})}}, {\sl
  \bibinfo{series}{Lecture Notes in Computer Science}} \bibinfo{volume}{10206},
  pp. \bibinfo{pages}{151--168}, \doi{10.1007/978-3-662-54580-5\_9}.

\bibitemdeclare{inproceedings}{BHKKPQTZ20}
\bibitem{BHKKPQTZ20}
\bibinfo{author}{Carlos~E. \surnamestart Budde\surnameend},
  \bibinfo{author}{Arnd \surnamestart Hartmanns\surnameend},
  \bibinfo{author}{Michaela \surnamestart Klauck\surnameend},
  \bibinfo{author}{Jan \surnamestart Kret{\'{\i}}nsk{\'{y}}\surnameend},
  \bibinfo{author}{David \surnamestart Parker\surnameend}, \bibinfo{author}{Tim
  \surnamestart Quatmann\surnameend}, \bibinfo{author}{Andrea \surnamestart
  Turrini\surnameend} \& \bibinfo{author}{Zhen \surnamestart Zhang\surnameend}
  (\bibinfo{year}{2020}): \emph{\bibinfo{title}{On Correctness, Precision, and
  Performance in Quantitative Verification ({QC}omp 2020 Competition Report)}}.
\newblock In \bibinfo{editor}{Tiziana \surnamestart Margaria\surnameend} \&
  \bibinfo{editor}{Bernhard \surnamestart Steffen\surnameend}, editors: {\sl
  \bibinfo{booktitle}{9th International Symposium on Leveraging Applications of
  Formal Methods ({ISoLA})}}, {\sl \bibinfo{series}{Lecture Notes in Computer
  Science}} \bibinfo{volume}{12479}, \bibinfo{publisher}{Springer}, pp.
  \bibinfo{pages}{216--241}, \doi{10.1007/978-3-030-83723-5\_15}.

\bibitemdeclare{article}{BHH21}
\bibitem{BHH21}
\bibinfo{author}{Yuliya \surnamestart Butkova\surnameend},
  \bibinfo{author}{Arnd \surnamestart Hartmanns\surnameend} \&
  \bibinfo{author}{Holger \surnamestart Hermanns\surnameend}
  (\bibinfo{year}{2021}): \emph{\bibinfo{title}{A {M}odest Approach to {M}arkov
  Automata}}.
\newblock {\sl \bibinfo{journal}{{ACM} Trans. Model. Comput. Simul.}}
  \bibinfo{volume}{31}(\bibinfo{number}{3}), pp. \bibinfo{pages}{14:1--14:34},
  \doi{10.1145/3449355}.

\bibitemdeclare{inproceedings}{DFH20}
\bibitem{DFH20}
\bibinfo{author}{Pedro~R. \surnamestart D'Argenio\surnameend},
  \bibinfo{author}{Juan~A. \surnamestart Fraire\surnameend} \&
  \bibinfo{author}{Arnd \surnamestart Hartmanns\surnameend}
  (\bibinfo{year}{2020}): \emph{\bibinfo{title}{Sampling Distributed Schedulers
  for Resilient Space Communication}}.
\newblock In \bibinfo{editor}{Ritchie \surnamestart Lee\surnameend},
  \bibinfo{editor}{Susmit \surnamestart Jha\surnameend} \&
  \bibinfo{editor}{Anastasia \surnamestart Mavridou\surnameend}, editors: {\sl
  \bibinfo{booktitle}{12th International {NASA} Formal Methods Symposium
  ({NFM})}}, {\sl \bibinfo{series}{Lecture Notes in Computer Science}}
  \bibinfo{volume}{12229}, \bibinfo{publisher}{Springer}, pp.
  \bibinfo{pages}{291--310}, \doi{10.1007/978-3-030-55754-6\_17}.

\bibitemdeclare{inproceedings}{DHLS16}
\bibitem{DHLS16}
\bibinfo{author}{Pedro~R. \surnamestart D'Argenio\surnameend},
  \bibinfo{author}{Arnd \surnamestart Hartmanns\surnameend},
  \bibinfo{author}{Axel \surnamestart Legay\surnameend} \&
  \bibinfo{author}{Sean \surnamestart Sedwards\surnameend}
  (\bibinfo{year}{2016}): \emph{\bibinfo{title}{Statistical Approximation of
  Optimal Schedulers for Probabilistic Timed Automata}}.
\newblock In \bibinfo{editor}{Erika \surnamestart
  {\'{A}}brah{\'{a}}m\surnameend} \& \bibinfo{editor}{Marieke \surnamestart
  Huisman\surnameend}, editors: {\sl \bibinfo{booktitle}{12th International
  Conference on Integrated Formal Methods ({iFM})}}, {\sl
  \bibinfo{series}{Lecture Notes in Computer Science}} \bibinfo{volume}{9681},
  \bibinfo{publisher}{Springer}, pp. \bibinfo{pages}{99--114},
  \doi{10.1007/978-3-319-33693-0\_7}.

\bibitemdeclare{inproceedings}{DHS18}
\bibitem{DHS18}
\bibinfo{author}{Pedro~R. \surnamestart D'Argenio\surnameend},
  \bibinfo{author}{Arnd \surnamestart Hartmanns\surnameend} \&
  \bibinfo{author}{Sean \surnamestart Sedwards\surnameend}
  (\bibinfo{year}{2018}): \emph{\bibinfo{title}{Lightweight Statistical Model
  Checking in Nondeterministic Continuous Time}}.
\newblock In \bibinfo{editor}{Tiziana \surnamestart Margaria\surnameend} \&
  \bibinfo{editor}{Bernhard \surnamestart Steffen\surnameend}, editors: {\sl
  \bibinfo{booktitle}{8th International Symposium on Leveraging Applications of
  Formal Methods, Verification and Validation ({ISoLA})}}, {\sl
  \bibinfo{series}{Lecture Notes in Computer Science}} \bibinfo{volume}{11245},
  \bibinfo{publisher}{Springer}, pp. \bibinfo{pages}{336--353},
  \doi{10.1007/978-3-030-03421-4\_22}.

\bibitemdeclare{inproceedings}{DJJL02}
\bibitem{DJJL02}
\bibinfo{author}{Pedro~R. \surnamestart D'Argenio\surnameend},
  \bibinfo{author}{Bertrand \surnamestart Jeannet\surnameend},
  \bibinfo{author}{Henrik~Ejersbo \surnamestart Jensen\surnameend} \&
  \bibinfo{author}{Kim~Guldstrand \surnamestart Larsen\surnameend}
  (\bibinfo{year}{2002}): \emph{\bibinfo{title}{Reduction and Refinement
  Strategies for Probabilistic Analysis}}.
\newblock In \bibinfo{editor}{Holger \surnamestart Hermanns\surnameend} \&
  \bibinfo{editor}{Roberto \surnamestart Segala\surnameend}, editors: {\sl
  \bibinfo{booktitle}{Second Joint International Workshop on Process Algebra
  and Probabilistic Methods, Performance Modeling and Verification
  ({PAPM-PROBMIV})}}, {\sl \bibinfo{series}{Lecture Notes in Computer Science}}
  \bibinfo{volume}{2399}, \bibinfo{publisher}{Springer}, pp.
  \bibinfo{pages}{57--76}, \doi{10.1007/3-540-45605-8\_5}.

\bibitemdeclare{inproceedings}{DLLMW11}
\bibitem{DLLMW11}
\bibinfo{author}{Alexandre \surnamestart David\surnameend},
  \bibinfo{author}{Kim~G. \surnamestart Larsen\surnameend},
  \bibinfo{author}{Axel \surnamestart Legay\surnameend},
  \bibinfo{author}{Marius \surnamestart Mikucionis\surnameend} \&
  \bibinfo{author}{Zheng \surnamestart Wang\surnameend} (\bibinfo{year}{2011}):
  \emph{\bibinfo{title}{Time for Statistical Model Checking of Real-Time
  Systems}}.
\newblock In \bibinfo{editor}{Ganesh \surnamestart Gopalakrishnan\surnameend}
  \& \bibinfo{editor}{Shaz \surnamestart Qadeer\surnameend}, editors: {\sl
  \bibinfo{booktitle}{23rd International Conference on Computer Aided
  Verification ({CAV})}}, {\sl \bibinfo{series}{Lecture Notes in Computer
  Science}} \bibinfo{volume}{6806}, \bibinfo{publisher}{Springer}, pp.
  \bibinfo{pages}{349--355}, \doi{10.1007/978-3-642-22110-1\_27}.

\bibitemdeclare{inproceedings}{DJKV17}
\bibitem{DJKV17}
\bibinfo{author}{Christian \surnamestart Dehnert\surnameend},
  \bibinfo{author}{Sebastian \surnamestart Junges\surnameend},
  \bibinfo{author}{Joost-Pieter \surnamestart Katoen\surnameend} \&
  \bibinfo{author}{Matthias \surnamestart Volk\surnameend}
  (\bibinfo{year}{2017}): \emph{\bibinfo{title}{A {S}torm is Coming: A Modern
  Probabilistic Model Checker}}.
\newblock In \bibinfo{editor}{Rupak \surnamestart Majumdar\surnameend} \&
  \bibinfo{editor}{Viktor \surnamestart Kuncak\surnameend}, editors: {\sl
  \bibinfo{booktitle}{29th International Conference on Computer Aided
  Verification ({CAV})}}, {\sl \bibinfo{series}{Lecture Notes in Computer
  Science}} \bibinfo{volume}{10427}, \bibinfo{publisher}{Springer}, pp.
  \bibinfo{pages}{592--600}, \doi{10.1007/978-3-319-63390-9\_31}.

\bibitemdeclare{inproceedings}{EHZ10}
\bibitem{EHZ10}
\bibinfo{author}{Christian \surnamestart Eisentraut\surnameend},
  \bibinfo{author}{Holger \surnamestart Hermanns\surnameend} \&
  \bibinfo{author}{Lijun \surnamestart Zhang\surnameend}
  (\bibinfo{year}{2010}): \emph{\bibinfo{title}{On Probabilistic Automata in
  Continuous Time}}.
\newblock In: {\sl \bibinfo{booktitle}{25th Annual {IEEE} Symposium on Logic in
  Computer Science ({LICS})}}, \bibinfo{publisher}{{IEEE} Computer Society},
  pp. \bibinfo{pages}{342--351}, \doi{10.1109/LICS.2010.41}.

\bibitemdeclare{inproceedings}{FC18}
\bibitem{FC18}
\bibinfo{author}{Ansgar \surnamestart Fehnker\surnameend} \&
  \bibinfo{author}{Kaylash \surnamestart Chaudhary\surnameend}
  (\bibinfo{year}{2018}): \emph{\bibinfo{title}{Twenty Percent and a Few Days
  -- Optimising a {B}itcoin Majority Attack}}.
\newblock In \bibinfo{editor}{Aaron \surnamestart Dutle\surnameend},
  \bibinfo{editor}{C{\'{e}}sar~A. \surnamestart Mu{\~{n}}oz\surnameend} \&
  \bibinfo{editor}{Anthony \surnamestart Narkawicz\surnameend}, editors: {\sl
  \bibinfo{booktitle}{10th International {NASA} Formal Methods Symposium
  ({NFM})}}, {\sl \bibinfo{series}{Lecture Notes in Computer Science}}
  \bibinfo{volume}{10811}, \bibinfo{publisher}{Springer}, pp.
  \bibinfo{pages}{157--163}, \doi{10.1007/978-3-319-77935-5\_11}.

\bibitemdeclare{inproceedings}{FHHWZ11}
\bibitem{FHHWZ11}
\bibinfo{author}{Martin \surnamestart Fr{\"{a}}nzle\surnameend},
  \bibinfo{author}{Ernst~Moritz \surnamestart Hahn\surnameend},
  \bibinfo{author}{Holger \surnamestart Hermanns\surnameend},
  \bibinfo{author}{Nicol{\'{a}}s \surnamestart Wolovick\surnameend} \&
  \bibinfo{author}{Lijun \surnamestart Zhang\surnameend}
  (\bibinfo{year}{2011}): \emph{\bibinfo{title}{Measurability and safety
  verification for stochastic hybrid systems}}.
\newblock In \bibinfo{editor}{Marco \surnamestart Caccamo\surnameend},
  \bibinfo{editor}{Emilio \surnamestart Frazzoli\surnameend} \&
  \bibinfo{editor}{Radu \surnamestart Grosu\surnameend}, editors: {\sl
  \bibinfo{booktitle}{14th {ACM} International Conference on Hybrid Systems:
  Computation and Control ({HSCC})}}, \bibinfo{publisher}{{ACM}}, pp.
  \bibinfo{pages}{43--52}, \doi{10.1145/1967701.1967710}.

\bibitemdeclare{inproceedings}{GD07}
\bibitem{GD07}
\bibinfo{author}{Sergio \surnamestart Giro\surnameend} \&
  \bibinfo{author}{Pedro~R. \surnamestart D'Argenio\surnameend}
  (\bibinfo{year}{2007}): \emph{\bibinfo{title}{Quantitative Model Checking
  Revisited: Neither Decidable Nor Approximable}}.
\newblock In \bibinfo{editor}{Jean-Fran{\c{c}}ois \surnamestart
  Raskin\surnameend} \& \bibinfo{editor}{P.~S. \surnamestart
  Thiagarajan\surnameend}, editors: {\sl \bibinfo{booktitle}{5th International
  Conference on Formal Modeling and Analysis of Timed Systems ({FORMATS})}},
  {\sl \bibinfo{series}{Lecture Notes in Computer Science}}
  \bibinfo{volume}{4763}, \bibinfo{publisher}{Springer}, pp.
  \bibinfo{pages}{179--194}, \doi{10.1007/978-3-540-75454-1\_14}.

\bibitemdeclare{article}{GD09}
\bibitem{GD09}
\bibinfo{author}{Sergio \surnamestart Giro\surnameend} \&
  \bibinfo{author}{Pedro~R. \surnamestart D'Argenio\surnameend}
  (\bibinfo{year}{2009}): \emph{\bibinfo{title}{On the Expressive Power of
  Schedulers in Distributed Probabilistic Systems}}.
\newblock {\sl \bibinfo{journal}{Electron. Notes Theor. Comput. Sci.}}
  \bibinfo{volume}{253}(\bibinfo{number}{3}), pp. \bibinfo{pages}{45--71},
  \doi{10.1016/j.entcs.2009.10.005}.

\bibitemdeclare{inproceedings}{HH16}
\bibitem{HH16}
\bibinfo{author}{Ernst~Moritz \surnamestart Hahn\surnameend} \&
  \bibinfo{author}{Arnd \surnamestart Hartmanns\surnameend}
  (\bibinfo{year}{2016}): \emph{\bibinfo{title}{A Comparison of Time- and
  Reward-Bounded Probabilistic Model Checking Techniques}}.
\newblock In \bibinfo{editor}{Martin \surnamestart Fr{\"{a}}nzle\surnameend},
  \bibinfo{editor}{Deepak \surnamestart Kapur\surnameend} \&
  \bibinfo{editor}{Naijun \surnamestart Zhan\surnameend}, editors: {\sl
  \bibinfo{booktitle}{Second International Symposium on Dependable Software
  Engineering: Theories, Tools, and Applications ({SETTA})}}, {\sl
  \bibinfo{series}{Lecture Notes in Computer Science}} \bibinfo{volume}{9984},
  pp. \bibinfo{pages}{85--100}, \doi{10.1007/978-3-319-47677-3\_6}.

\bibitemdeclare{inproceedings}{HHHKKKPQRS19}
\bibitem{HHHKKKPQRS19}
\bibinfo{author}{Ernst~Moritz \surnamestart Hahn\surnameend},
  \bibinfo{author}{Arnd \surnamestart Hartmanns\surnameend},
  \bibinfo{author}{Christian \surnamestart Hensel\surnameend},
  \bibinfo{author}{Michaela \surnamestart Klauck\surnameend},
  \bibinfo{author}{Joachim \surnamestart Klein\surnameend},
  \bibinfo{author}{Jan \surnamestart Kret{\'{\i}}nsk{\'{y}}\surnameend},
  \bibinfo{author}{David \surnamestart Parker\surnameend}, \bibinfo{author}{Tim
  \surnamestart Quatmann\surnameend}, \bibinfo{author}{Enno \surnamestart
  Ruijters\surnameend} \& \bibinfo{author}{Marcel \surnamestart
  Steinmetz\surnameend} (\bibinfo{year}{2019}): \emph{\bibinfo{title}{The 2019
  Comparison of Tools for the Analysis of Quantitative Formal Models ({QC}omp
  2019 Competition Report)}}.
\newblock In \bibinfo{editor}{Dirk \surnamestart Beyer\surnameend},
  \bibinfo{editor}{Marieke \surnamestart Huisman\surnameend},
  \bibinfo{editor}{Fabrice \surnamestart Kordon\surnameend} \&
  \bibinfo{editor}{Bernhard \surnamestart Steffen\surnameend}, editors: {\sl
  \bibinfo{booktitle}{25 Years of {TACAS}: {TOOL}ympics}}, {\sl
  \bibinfo{series}{Lecture Notes in Computer Science}} \bibinfo{volume}{11429},
  \bibinfo{publisher}{Springer}, pp. \bibinfo{pages}{69--92},
  \doi{10.1007/978-3-030-17502-3\_5}.

\bibitemdeclare{article}{HHHK13}
\bibitem{HHHK13}
\bibinfo{author}{Ernst~Moritz \surnamestart Hahn\surnameend},
  \bibinfo{author}{Arnd \surnamestart Hartmanns\surnameend},
  \bibinfo{author}{Holger \surnamestart Hermanns\surnameend} \&
  \bibinfo{author}{Joost-Pieter \surnamestart Katoen\surnameend}
  (\bibinfo{year}{2013}): \emph{\bibinfo{title}{A compositional modelling and
  analysis framework for stochastic hybrid systems}}.
\newblock {\sl \bibinfo{journal}{Formal Methods Syst. Des.}}
  \bibinfo{volume}{43}(\bibinfo{number}{2}), pp. \bibinfo{pages}{191--232},
  \doi{10.1007/s10703-012-0167-z}.

\bibitemdeclare{inproceedings}{HH14}
\bibitem{HH14}
\bibinfo{author}{Arnd \surnamestart Hartmanns\surnameend} \&
  \bibinfo{author}{Holger \surnamestart Hermanns\surnameend}
  (\bibinfo{year}{2014}): \emph{\bibinfo{title}{The {M}odest {T}oolset: An
  Integrated Environment for Quantitative Modelling and Verification}}.
\newblock In \bibinfo{editor}{Erika \surnamestart
  {\'{A}}brah{\'{a}}m\surnameend} \& \bibinfo{editor}{Klaus \surnamestart
  Havelund\surnameend}, editors: {\sl \bibinfo{booktitle}{20th International
  Conference on Tools and Algorithms for the Construction and Analysis of
  Systems ({TACAS})}}, {\sl \bibinfo{series}{Lecture Notes in Computer
  Science}} \bibinfo{volume}{8413}, \bibinfo{publisher}{Springer}, pp.
  \bibinfo{pages}{593--598}, \doi{10.1007/978-3-642-54862-8\_51}.

\bibitemdeclare{inproceedings}{HH15}
\bibitem{HH15}
\bibinfo{author}{Arnd \surnamestart Hartmanns\surnameend} \&
  \bibinfo{author}{Holger \surnamestart Hermanns\surnameend}
  (\bibinfo{year}{2015}): \emph{\bibinfo{title}{Explicit Model Checking of Very
  Large {MDP} Using Partitioning and Secondary Storage}}.
\newblock In \bibinfo{editor}{Bernd \surnamestart Finkbeiner\surnameend},
  \bibinfo{editor}{Geguang \surnamestart Pu\surnameend} \&
  \bibinfo{editor}{Lijun \surnamestart Zhang\surnameend}, editors: {\sl
  \bibinfo{booktitle}{13th International Symposium on Automated Technology for
  Verification and Analysis ({ATVA})}}, {\sl \bibinfo{series}{Lecture Notes in
  Computer Science}} \bibinfo{volume}{9364}, \bibinfo{publisher}{Springer}, pp.
  \bibinfo{pages}{131--147}, \doi{10.1007/978-3-319-24953-7\_10}.

\bibitemdeclare{inproceedings}{HH19}
\bibitem{HH19}
\bibinfo{author}{Arnd \surnamestart Hartmanns\surnameend} \&
  \bibinfo{author}{Holger \surnamestart Hermanns\surnameend}
  (\bibinfo{year}{2019}): \emph{\bibinfo{title}{A {M}odest {M}arkov Automata
  Tutorial}}.
\newblock In \bibinfo{editor}{Markus \surnamestart Kr{\"{o}}tzsch\surnameend}
  \& \bibinfo{editor}{Daria \surnamestart Stepanova\surnameend}, editors: {\sl
  \bibinfo{booktitle}{15th International Reasoning Web Summer School}}, {\sl
  \bibinfo{series}{Lecture Notes in Computer Science}} \bibinfo{volume}{11810},
  \bibinfo{publisher}{Springer}, pp. \bibinfo{pages}{250--276},
  \doi{10.1007/978-3-030-31423-1\_8}.

\bibitemdeclare{inproceedings}{HKPQR19}
\bibitem{HKPQR19}
\bibinfo{author}{Arnd \surnamestart Hartmanns\surnameend},
  \bibinfo{author}{Michaela \surnamestart Klauck\surnameend},
  \bibinfo{author}{David \surnamestart Parker\surnameend}, \bibinfo{author}{Tim
  \surnamestart Quatmann\surnameend} \& \bibinfo{author}{Enno \surnamestart
  Ruijters\surnameend} (\bibinfo{year}{2019}): \emph{\bibinfo{title}{The
  Quantitative Verification Benchmark Set}}.
\newblock In \bibinfo{editor}{Tom{\'{a}}s \surnamestart Vojnar\surnameend} \&
  \bibinfo{editor}{Lijun \surnamestart Zhang\surnameend}, editors: {\sl
  \bibinfo{booktitle}{25th International Conference on Tools and Algorithms for
  the Construction and Analysis of Systems ({TACAS})}}, {\sl
  \bibinfo{series}{Lecture Notes in Computer Science}} \bibinfo{volume}{11427},
  \bibinfo{publisher}{Springer}, pp. \bibinfo{pages}{344--350},
  \doi{10.1007/978-3-030-17462-0\_20}.

\bibitemdeclare{inproceedings}{HSD17}
\bibitem{HSD17}
\bibinfo{author}{Arnd \surnamestart Hartmanns\surnameend},
  \bibinfo{author}{Sean \surnamestart Sedwards\surnameend} \&
  \bibinfo{author}{Pedro~R. \surnamestart D'Argenio\surnameend}
  (\bibinfo{year}{2017}): \emph{\bibinfo{title}{Efficient simulation-based
  verification of probabilistic timed automata}}.
\newblock In: {\sl \bibinfo{booktitle}{2017 Winter Simulation Conference
  ({WSC})}}, \bibinfo{publisher}{{IEEE}}, pp. \bibinfo{pages}{1419--1430},
  \doi{10.1109/WSC.2017.8247885}.

\bibitemdeclare{book}{How60}
\bibitem{How60}
\bibinfo{author}{Ronald~A. \surnamestart Howard\surnameend}
  (\bibinfo{year}{1960}): \emph{\bibinfo{title}{Dynamic Programming and
  {M}arkov Processes}}.
\newblock \bibinfo{publisher}{MIT Press}.

\bibitemdeclare{inproceedings}{KH21}
\bibitem{KH21}
\bibinfo{author}{Michaela \surnamestart Klauck\surnameend} \&
  \bibinfo{author}{Holger \surnamestart Hermanns\surnameend}
  (\bibinfo{year}{2021}): \emph{\bibinfo{title}{A {M}odest Approach to Dynamic
  Heuristic Search in Probabilistic Model Checking}}.
\newblock In \bibinfo{editor}{Alessandro \surnamestart Abate\surnameend} \&
  \bibinfo{editor}{Andrea \surnamestart Marin\surnameend}, editors: {\sl
  \bibinfo{booktitle}{18th International Conference on Quantitative Evaluation
  of Systems ({QEST})}}, {\sl \bibinfo{series}{Lecture Notes in Computer
  Science}} \bibinfo{volume}{12846}, \bibinfo{publisher}{Springer}, pp.
  \bibinfo{pages}{15--38}, \doi{10.1007/978-3-030-85172-9\_2}.

\bibitemdeclare{inproceedings}{KKH21}
\bibitem{KKH21}
\bibinfo{author}{Maximilian~A. \surnamestart K{\"{o}}hl\surnameend},
  \bibinfo{author}{Michaela \surnamestart Klauck\surnameend} \&
  \bibinfo{author}{Holger \surnamestart Hermanns\surnameend}
  (\bibinfo{year}{2021}): \emph{\bibinfo{title}{{M}omba: {JANI} Meets
  {P}ython}}.
\newblock In \bibinfo{editor}{Jan~Friso \surnamestart Groote\surnameend} \&
  \bibinfo{editor}{Kim~Guldstrand \surnamestart Larsen\surnameend}, editors:
  {\sl \bibinfo{booktitle}{27th International Conference on Tools and
  Algorithms for the Construction and Analysis of Systems ({TACAS})}}, {\sl
  \bibinfo{series}{Lecture Notes in Computer Science}} \bibinfo{volume}{12652},
  \bibinfo{publisher}{Springer}, pp. \bibinfo{pages}{389--398},
  \doi{10.1007/978-3-030-72013-1\_23}.

\bibitemdeclare{inproceedings}{KNP11}
\bibitem{KNP11}
\bibinfo{author}{Marta~Z. \surnamestart Kwiatkowska\surnameend},
  \bibinfo{author}{Gethin \surnamestart Norman\surnameend} \&
  \bibinfo{author}{David \surnamestart Parker\surnameend}
  (\bibinfo{year}{2011}): \emph{\bibinfo{title}{{PRISM} 4.0: Verification of
  Probabilistic Real-Time Systems}}.
\newblock In \bibinfo{editor}{Ganesh \surnamestart Gopalakrishnan\surnameend}
  \& \bibinfo{editor}{Shaz \surnamestart Qadeer\surnameend}, editors: {\sl
  \bibinfo{booktitle}{23rd International Conference on Computer Aided
  Verification ({CAV})}}, {\sl \bibinfo{series}{Lecture Notes in Computer
  Science}} \bibinfo{volume}{6806}, \bibinfo{publisher}{Springer}, pp.
  \bibinfo{pages}{585--591}, \doi{10.1007/978-3-642-22110-1\_47}.

\bibitemdeclare{article}{KNPS06}
\bibitem{KNPS06}
\bibinfo{author}{Marta~Z. \surnamestart Kwiatkowska\surnameend},
  \bibinfo{author}{Gethin \surnamestart Norman\surnameend},
  \bibinfo{author}{David \surnamestart Parker\surnameend} \&
  \bibinfo{author}{Jeremy \surnamestart Sproston\surnameend}
  (\bibinfo{year}{2006}): \emph{\bibinfo{title}{Performance analysis of
  probabilistic timed automata using digital clocks}}.
\newblock {\sl \bibinfo{journal}{Formal Methods Syst. Des.}}
  \bibinfo{volume}{29}(\bibinfo{number}{1}), pp. \bibinfo{pages}{33--78},
  \doi{10.1007/s10703-006-0005-2}.

\bibitemdeclare{article}{KNSS02}
\bibitem{KNSS02}
\bibinfo{author}{Marta~Z. \surnamestart Kwiatkowska\surnameend},
  \bibinfo{author}{Gethin \surnamestart Norman\surnameend},
  \bibinfo{author}{Roberto \surnamestart Segala\surnameend} \&
  \bibinfo{author}{Jeremy \surnamestart Sproston\surnameend}
  (\bibinfo{year}{2002}): \emph{\bibinfo{title}{Automatic verification of
  real-time systems with discrete probability distributions}}.
\newblock {\sl \bibinfo{journal}{Theor. Comput. Sci.}}
  \bibinfo{volume}{282}(\bibinfo{number}{1}), pp. \bibinfo{pages}{101--150},
  \doi{10.1016/S0304-3975(01)00046-9}.

\bibitemdeclare{inproceedings}{LST14}
\bibitem{LST14}
\bibinfo{author}{Axel \surnamestart Legay\surnameend}, \bibinfo{author}{Sean
  \surnamestart Sedwards\surnameend} \& \bibinfo{author}{Louis-Marie
  \surnamestart Traonouez\surnameend} (\bibinfo{year}{2014}):
  \emph{\bibinfo{title}{Scalable Verification of {M}arkov Decision Processes}}.
\newblock In \bibinfo{editor}{Carlos \surnamestart Canal\surnameend} \&
  \bibinfo{editor}{Akram \surnamestart Idani\surnameend}, editors: {\sl
  \bibinfo{booktitle}{4th Workshop on Formal Methods in the Development of
  Software ({WS-FMDS})}}, {\sl \bibinfo{series}{Lecture Notes in Computer
  Science}} \bibinfo{volume}{8938}, \bibinfo{publisher}{Springer}, pp.
  \bibinfo{pages}{350--362}, \doi{10.1007/978-3-319-15201-1\_23}.

\bibitemdeclare{inproceedings}{LHBSCRZ19}
\bibitem{LHBSCRZ19}
\bibinfo{author}{Benjamin \surnamestart Lewis\surnameend},
  \bibinfo{author}{Arnd \surnamestart Hartmanns\surnameend},
  \bibinfo{author}{Prabal \surnamestart Basu\surnameend},
  \bibinfo{author}{Rajesh~Jayashankara \surnamestart Shridevi\surnameend},
  \bibinfo{author}{Koushik \surnamestart Chakraborty\surnameend},
  \bibinfo{author}{Sanghamitra \surnamestart Roy\surnameend} \&
  \bibinfo{author}{Zhen \surnamestart Zhang\surnameend} (\bibinfo{year}{2019}):
  \emph{\bibinfo{title}{Probabilistic Verification for Reliable Network-on-Chip
  System Design}}.
\newblock In \bibinfo{editor}{Kim~Guldstrand \surnamestart Larsen\surnameend}
  \& \bibinfo{editor}{Tim A.~C. \surnamestart Willemse\surnameend}, editors:
  {\sl \bibinfo{booktitle}{24th International Conference on Formal Methods for
  Industrial Critical Systems ({FMICS})}}, {\sl \bibinfo{series}{Lecture Notes
  in Computer Science}} \bibinfo{volume}{11687}, \bibinfo{publisher}{Springer},
  pp. \bibinfo{pages}{110--126}, \doi{10.1007/978-3-030-27008-7\_7}.

\bibitemdeclare{article}{RFMDFD21}
\bibitem{RFMDFD21}
\bibinfo{author}{Fernando~D. \surnamestart Raverta\surnameend},
  \bibinfo{author}{Juan~A. \surnamestart Fraire\surnameend},
  \bibinfo{author}{Pablo~G. \surnamestart Madoery\surnameend},
  \bibinfo{author}{Ramiro~A. \surnamestart Demasi\surnameend},
  \bibinfo{author}{Jorge~M. \surnamestart Finochietto\surnameend} \&
  \bibinfo{author}{Pedro~R. \surnamestart D'Argenio\surnameend}
  (\bibinfo{year}{2021}): \emph{\bibinfo{title}{Routing in Delay-Tolerant
  Networks under uncertain contact plans}}.
\newblock {\sl \bibinfo{journal}{Ad Hoc Networks}} \bibinfo{volume}{123}, p.
  \bibinfo{pages}{102663}, \doi{10.1016/j.adhoc.2021.102663}.

\bibitemdeclare{inproceedings}{RLHBRCZ21}
\bibitem{RLHBRCZ21}
\bibinfo{author}{Riley \surnamestart Roberts\surnameend},
  \bibinfo{author}{Benjamin \surnamestart Lewis\surnameend},
  \bibinfo{author}{Arnd \surnamestart Hartmanns\surnameend},
  \bibinfo{author}{Prabal \surnamestart Basu\surnameend},
  \bibinfo{author}{Sanghamitra \surnamestart Roy\surnameend},
  \bibinfo{author}{Koushik \surnamestart Chakraborty\surnameend} \&
  \bibinfo{author}{Zhen \surnamestart Zhang\surnameend} (\bibinfo{year}{2021}):
  \emph{\bibinfo{title}{Probabilistic Verification for Reliability of a
  Two-by-Two Network-on-Chip System}}.
\newblock In \bibinfo{editor}{Alberto \surnamestart Lluch-Lafuente\surnameend}
  \& \bibinfo{editor}{Anastasia \surnamestart Mavridou\surnameend}, editors:
  {\sl \bibinfo{booktitle}{26th International Conference on Formal Methods for
  Industrial Critical Systems ({FMICS})}}, {\sl \bibinfo{series}{Lecture Notes
  in Computer Science}} \bibinfo{volume}{12863}, \bibinfo{publisher}{Springer},
  pp. \bibinfo{pages}{232--248}, \doi{10.1007/978-3-030-85248-1\_16}.

\bibitemdeclare{book}{RT09}
\bibitem{RT09}
\bibinfo{editor}{Gerardo \surnamestart Rubino\surnameend} \&
  \bibinfo{editor}{Bruno \surnamestart Tuffin\surnameend}, editors
  (\bibinfo{year}{2009}): \emph{\bibinfo{title}{Rare Event Simulation using
  {M}onte {C}arlo Methods}}.
\newblock \bibinfo{publisher}{Wiley}, \doi{10.1002/9780470745403}.

\bibitemdeclare{inproceedings}{Spr00}
\bibitem{Spr00}
\bibinfo{author}{Jeremy \surnamestart Sproston\surnameend}
  (\bibinfo{year}{2000}): \emph{\bibinfo{title}{Decidable Model Checking of
  Probabilistic Hybrid Automata}}.
\newblock In \bibinfo{editor}{Mathai \surnamestart Joseph\surnameend}, editor:
  {\sl \bibinfo{booktitle}{6th International Symposium on Formal Techniques in
  Real-Time and Fault-Tolerant Systems ({FTRTFT})}}, {\sl
  \bibinfo{series}{Lecture Notes in Computer Science}} \bibinfo{volume}{1926},
  \bibinfo{publisher}{Springer}, pp. \bibinfo{pages}{31--45},
  \doi{10.1007/3-540-45352-0\_5}.

\end{thebibliography}
\end{document}